\RequirePackage[2020-02-02]{latexrelease}
\documentclass[aps,pra,floatfix,amsmath,amssymb,superscriptaddress]{revtex4}

\usepackage{latexsym}
\usepackage{epsfig}
\usepackage{dcolumn}
\usepackage{amsmath,bm}
\usepackage{amssymb}
\usepackage{subcaption}
\usepackage{graphicx}
\usepackage{diagbox}
\usepackage{times}
\usepackage[section]{placeins}
\usepackage{booktabs}
\usepackage{color}
\usepackage{float}

\def\be{\begin{equation}}
\def\ee{\end{equation}}

\begin{document}
	
	\title{Effective detection of quantum discord by using Convolutional Neural Networks}
\vspace{2cm}
\author{N. Taghadomi}
\affiliation{Department of Physics, Tarbiat Modares University, Tehran, Iran.}
\author{A. Mani \footnote{email: mani.azam@ut.ac.ir}}
\affiliation{Department of Engineering Science, College of Engineering, University of Tehran, Iran}
\author{A. Fahim}
\affiliation{Department of Engineering Science, College of Engineering, University of Tehran, Iran}
\author{A. Bakouei}
\affiliation{Department of Physics, Tarbiat Modares University, Tehran, Iran.}	
\author{M. S. Salami}
\affiliation{Department of Computer Engineering, Sharif University of Technology, Tehran, Iran}

\vspace{2cm}	

	\begin{abstract}
		Quantum discord is a form of correlation that is defined as the difference between quantum and classical mutual information of two parties. Due to the optimization involved in the definition of classical mutual information of quantum systems, calculating and distinguishing between discordant and non-discordant states is not a trivial task. Additionally, complete tomography of a quantum state is the prerequisite for the calculation of its quantum discord, and it is indeed resource-consuming.
		Here, by using the relation between the kernels of the convolutional layers of an artificial neural network and the expectation value of operators in quantum mechanical measurements, we design a Convolutional Neural Network (CNN) that uses $16$ kernels to completely distinguish between the discordant and non-discordant general two-qubit states. We have also designed a Branching Convolutional Neural Network (BCNN) that can effectively detect quantum discord. Our BCNN achieves an accuracy of around $85\%$ or $99\%$, by utilizing only $5$ or $8$ kernels, respectively. Our results show that to detect the existence of quantum discord up to the desired accuracy, instead of complete tomography, one can use suitable quantum circuits to directly measure the expectation values of the kernels, and then a fully connected network will solve the detection problem. 
    	\end{abstract}

\pacs{03.67.-a ,03.65.-w }
	
	\date{\today}
	
	\maketitle
\vspace{0cm}

\section{Introduction}

Quantum Discord (QD), as one of the essential resources in quantum information theory, is a  topic of current interest in various fields such as quantum teleportation \cite{QD teleportation}, quantum key distribution \cite{Quantum discord in cryptography}, quantum thermodynamics \cite{Quantum discord in Thermodynamic1, Quantum discord in Thermodynamic2}, quantum computing \cite{Quantum discord in Quantum Computing}, and  quantum metrology \cite{Quantum discord in metrology1, Quantum discord in metrology2}. QD is defined as the difference between the total correlations of two parties and their classical correlations. Even in the case of complete access to the quantum state (complete quantum tomography), general detection and calculation of quantum discord is challenging and time-consuming.  In fact, the problem belongs to the NP-hard class \cite{NP_discord}, and its difficulty relates to the optimization which lies in the definition of classical correlations. It should be noted that there are some limited works that report a closed form for QD of special classes of quantum states, such as  two-qubit X-states \cite{Luo2008,Ali2010,Girolami2011,Chen2011,Maldonado2015, Quantum discord in tomography, closed2}, but there is not a general closed formula for DQ and most of the existing formulae are valid with some small values of error \cite{Huang}, even for the smallest two-partite systems, i.e. general two-qubit states.\\

Over the last few years, machine learning methods have permeated quantum science and its applications, such as in quantum state tomography \cite{NNtomography, state tomography CNN}, detection and classification of quantum correlations \cite{raeisi, Chen 2022, Asif2023, Roik 2022, Discord network, discord algorithm, entanglementClassification1, entanglementClassification2}, quantum error correction \cite{error correction net.}, and design of quantum circuits \cite{quantum circuit ML}. In particular, the possibility of finding high-performance entanglement witnesses that do not require complete tomography was shown in \cite{raeisi, greenwood}, and the technique of supervised learning was utilized in \cite{Chen 2022, Asif2023} for entanglement detection and classification. Besides the detection and witnessing problems, there are also some works that use machine learning to predict and estimate the amount of the entanglement of quantum states \cite{Roik 2022}. Similar to entanglement, machine learning methods are gaining popularity as the means for the detection and calculation of quantum discord. For example, the quantum discord of X- and real two-qubit states are studied by a neural network with prior knowledge \cite{Discord network}, and also a quantum computing algorithm has been proposed for the calculation of quantum discord of two-qubit systems \cite{discord algorithm}.\\

In addition to ordinary neural networks that only consist of fully connected layers, there are other types of neural networks that are constructed from a sequence of layers that apply filters (kernels) to the input data in order to extract the efficient parameters of the problem. These networks are inspired by the natural visual perception mechanisms observed in living creatures, and they use a deep learning architecture known as convolutional neural network (CNN) \cite{Hubel and Wiesel’s research,origin of the CNN architecture,Introduce CNN}. Primarily, CNNs were  used for image-processing tasks such as image recognition, but in recent times, they have found applications in various other scientific fields, including chemistry \cite{CNN in chemistry}, medical image analysis \cite{CNN in medical image analysis}, and also quantum physics \cite{state tomography CNN, phase detection, NISQ devices NN, efficient observable operators, entanglement with Siamese}. Explicitly,  the CNN method has been shown to be useful in many quantum problems, such as efficient quantum state tomography \cite{state tomography CNN}, determining the phase diagram \cite{phase detection}, reducing the computing complexity in the noisy intermediate-scale quantum (NISQ) era \cite{NISQ devices NN}, and also quantifying the entanglement \cite{efficient observable operators, entanglement with Siamese}.\\

In this article, by using convolutional neural networks, we propose a novel approach to effectively distinguish between general two-qubit states that have or do not have quantum discord (discordant or non-discordant states). 
We use observable operators as the kernels of our convolutional network. The outputs of the convolutional layers - which are the expectation values of the observable operators - are then used as the inputs of a fully connected layer in order to determine whether a quantum state has discord or not. Furthermore, we upgraded our CNN and designed a Branching Convolution Neural Network (BCNN), which its kernels act on the input through independent paths \cite{BCNN}. We show that our branching neural network can detect discordant states by using fewer kernels than CNN. In fact, one can classify discordant and non-discordant states with an accuracy of around $99\%$ by using $8$ kernels, and with a accuracy of $85\%$ by applying only $5$ kernels. Since the kernels correspond to observable operators, fewer kernels means fewer number of parameters that are necessary for the quantum discord detection problem with the desired accuracy. Based on these fewer kernels, we discuss that one can use quantum circuits to directly measure the necessary parameters (Not complete tomography of the state) and then use a fully connected network to decide the existence of quantum discord. Briefly stated, the final fully connected network seems to act as a quantum discord witness, which does not require complete tomography to decide whether a state is discordant or not. Another key advantage of our work is that our networks are designed to work for any general two-qubit state, and our study is not restricted to a special class of these states.  \\

The paper is organized as follows: In the next section, we present the required preliminaries by reviewing the definition of quantum discord and the criteria of zero discord. We will then explain the structure of our CNN and BCNN networks in section \ref{Machine learning model}. The results and discussions about our scheme are followed in section \ref{result}, and finally, the paper will end with a conclusion in section \ref{con.}.

\section{Preliminaries}
The correlation between two or more quantum systems can exhibit peculiar non-classical features that are not seen in classical systems. These features are known as quantum (non-classical) correlations. Quantum discord is a form of quantum correlation, which is defined as the difference between quantum and classical mutual information of two parties. It is worth mentioning that QD is more than entanglement, i.e. there are some separable states that have non-zero QD \cite{Ollivier,Henderson}.\\

The definition of quantum discord arises from the fact that two classically identical expressions of mutual information are no longer identical in the quantum regime. Consider a system of two parts, say A and B; the first expression for the mutual information, i.e. $I(A,B)$, which shows the total correlations of A and B, is
\be\label{mutual1}
I(\rho_{AB})= S(\rho_A)+S(\rho_B)-S(\rho_{AB}),
\ee
where $\rho_{AB}$ is the joint density matrix of the system, $\rho_{A}$ ($\rho_B$) is the reduced  density matrix of subsystem A (B), and $S(\bullet)$ shows the Von Neumann entropy. The second expression of mutual information, i.e. $J(A,B)$, which is regarded as the classical correlations of subsystems, is
\be\label{mutual2}
J(A|B)= S(\rho_A)-\min_{\Pi_k} S(\rho_{A|B}),
\ee
where the minimization is performed over all positive operator-valued measurements (POVM) of subsystem B $\{ \Pi_k=I^A\otimes \Pi_k^B\}$,  $S(\rho_{A|B})= \sum_k p_k S(\rho_k^{A})$ is the average conditional entropy of subsystem A after measuring B,  $p_k= Tr(\Pi_k\rho_{AB}\Pi_k^\dag$) is the probability of obtaining outcome $k$ and $\rho_{A|k}=Tr_B (\Pi_k\rho_{AB}\Pi_k^\dag) /p_k$.
\\

\noindent The formal expression of QD of $\rho_{AB}$ is then given by:
\be\label{Discord}
QD(\rho_{AB}) := I(\rho_{AB})-J(A|B)= S(\rho_B)-S(\rho_{AB})+ \min_{\Pi_k} S(\rho_{A|B}),
\ee
which is evidently not symmetric with respect to subsystems A and B. 
It has been shown that the above expression is equivalent to the case where the minimization is only performed over the rank one projective measurements \cite{Henderson, criteria non-discord}. If one considers a two-qubit system, the orthogonal measurements of subsystem B can be written as $\Pi^B_\pm=|n_\pm^B \rangle \langle n_\pm^B|$, where $|n_+^B\rangle=\cos\theta|0\rangle +e^{i\phi} \sin\theta |1\rangle$ and $|n_-^B\rangle=\sin\theta|0\rangle - e^{i\phi}\cos\theta |1\rangle$ are two orthogonal pure states. The minimization which is present in equation (\ref{Discord}), should then be performed over all possible values of $\theta$ and $\phi$. It should be noted that the states with zero quantum discord are the ones that do not change after a particular local measurement, i.e. for a non-discordant state there exist particular orthogonal states  $|n_\pm^B\rangle$, such that
\be\label{non-discord}
\begin{split}
\rho_{ND}&=\sum_i p_i \rho_i^A \otimes |n_i^B\rangle \langle n_i^B|\\
                     &=p_+ \rho^A_+ \otimes |n_+^B\rangle \langle n_+^B| + (1-p_+)\rho^A_- \otimes |n^B_-\rangle \langle n^B_-|,
\end{split}
\ee
where $0\leq p_+ \leq 1$, and $\rho^A_\pm$ are two $2 \times 2$ density matrices of subsystem A. Having in mind the general form (\ref{non-discord}) of non-discordant states, we will also use a criterion that is presented in \cite{criteria non-discord} for zero discord states. A general two-qubit state can be written in the form 
\be
\begin{split}
\rho_{AB}&=\sum_{i,j=0}^1|i\rangle \langle j|\otimes \mathcal{B}_{ij},\\
\end{split}
\ee
where its block representation is
\begin{equation}\label{non_criteria}
\rho_{AB}=
 \renewcommand{\arraystretch}{1.4}
  \left[\,\:\begin{matrix}
     \mathcal{B}_{11} & \vrule & \mathcal{B}_{12}  \\
    \hline
     \mathcal{B}_{21} & \vrule & \mathcal{B}_{22}  \\
    \end{matrix}
\,\: \right],
\end{equation}
then, it is known that the following two conditions are necessary and sufficient for the state $\rho_{AB}$ to have zero quantum discord \cite{criteria non-discord}
$$ [\mathcal{B}_{ij},\mathcal{B}_{ij}^\dagger]=0, \hspace{1cm} \forall \  i,j$$
$$ [\mathcal{B}_{ij},\mathcal{B}_{kl}]=0, \hspace{1cm} \ \forall i,j,k,l.$$
In view of the above equations, one can check whether a two-qubit state has quantum discord or not, but this is indeed the case if the complete tomography of the state is at disposal. Following, we will use a data set of discordant and non-discordant states to design a network that predicts the existence of quantum discord in any arbitrary state without having access to its complete tomography.

\section{Methods}\label{Machine learning model}

Consider a general two-qubit state;
\be\label{rho}
\rho=\frac{1}{4}\sum_{i,j=0}^3 C_{ij}\sigma_i \otimes \sigma_j ,
\ee
where $\sigma_0$ is the identity operator, $\sigma_{i\neq 0}$ are Pauli matrices, and $C_{ij}=Tr(\rho(\sigma_i \otimes \sigma_j))$ are  real coefficients.
Perfect tomography of the state is equivalent to complete access to all 15 independent coefficients $C_{ij}$, and it requires at least 15 measurements.
Below, we will briefly describe the structures of CNN and BCNN networks that we use with the aim of reducing the number of measurements that are required for the decision problem of quantum discord. In fact, these neural networks perform like the witnesses of QD, i.e. they take the quantum state as their input and then decide whether it has QD or not, i.e they return the number $0$ for non-discordant states and the number $1$ for the discordant states.\\

\begin{figure}[H]
	\vspace{-0.25cm}
	\includegraphics[scale=0.55]{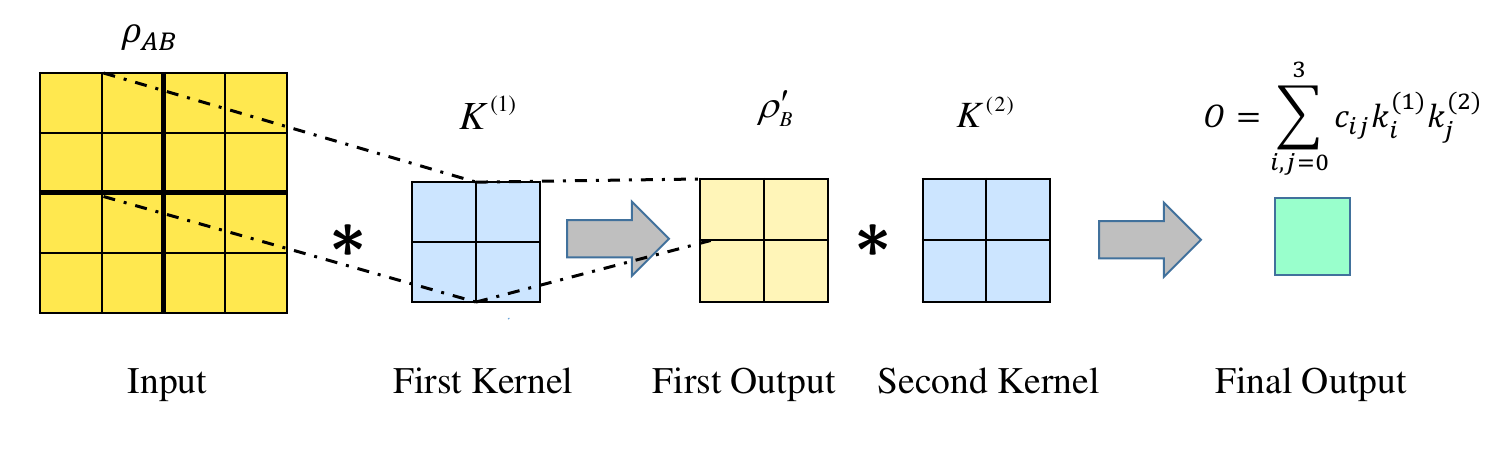}
	\centering
	\caption{Illustration of the convolution process: The $4\times 4$ input density matrix $\rho_{AB}$ passes through two convolutional layers, which are described by $2\times 2$ kernel matrices  ${{K}^{(1)}}$ and ${{K}^{(2)}}$. Convolution is the summation of element-wise multiplications. The output of the first layer is regarded as the input of the second one. The final output is shown with the number $O$. }
	\label{CNNsigma}
\end{figure}

In Convolutional Neural Networks, the positional information of the input data plays a crucial role in determining the output based on its neighboring values.
A CNN consists of some convolutional layers; through each, the efficient extraction of features is performed by applying the convolution operation on the input and the filtering matrices, known as kernels, see Figure \ref{CNNsigma}.
After the extraction of efficient features, a fully connected network is usually used to obtain the desired output.

Here, we apply a  two-layer CNN on the two-qubit state (\ref{rho}). In each layer, the elements of $2\times 2$ kernel matrices are multiplied by their corresponding elements of the input matrix, and the results add up to produce an output, which is regarded as an input item of the next layer. Suppose that kernels of the first and second layers are shown by ${{K}^{(1)}}$ and ${{K}^{(2)}}$, respectively. One can generally expand these matrices with respect to the conjugation of Pauli matrices, i.e.
\be\label{kernels1}
\begin{split}
  & {{K}^{(1)}}=\sum\limits_{n=0}^{3}{k_{n}^{(1)}}\sigma _{n}^{*}, \\ 
  & {{K}^{(2)}}=\sum\limits_{n=0}^{3}{k_{n}^{(2)}}\sigma _{n}^{*}. \\
\end{split}
\ee
Note that here, we have used the conjugation of Pauli matrices (instead of Pauli matrices themselves) for the sake of future simplicity of notation. After applying the first kernel, ${{K}^{(1)}}$, by using the convolutional relation ${{\sigma }_{i}}*\sigma _{j}^{*}=2{{\delta }_{ij}}$, the resulting contracted matrix $\rho'_B$ will be:\\
\be \label{k1out}
\begin{split}
   \rho'_B&=\frac{1}{4}\sum\limits_{i,j=0}^{3}{{{c}_{ij}}}({{\sigma }_{i}}*{{K}^{(1)}})\otimes {{\sigma }_{j}}, \\  
    & =\frac{1}{4}\sum\limits_{i,j,n=0}^{3}{{c}_{ij}}{k_{n}^{(1)}} (\sigma_i * \sigma_n^*) \ {{\sigma }_{j}}\\
    &=\frac{1}{2}\sum\limits_{i,j=0}^{3}{k_{i}^{(1)}{{c}_{ij}}}{{\sigma }_{j}}.
\end{split}
\ee
The ultimate convolution output, which is a single number $O$, will be obtained by applying the second kernel $K^{(2)}$ on (\ref{k1out}) after setting the stride equal to $2$,
\be \label{k2out}
\begin{split}
   O&=\frac{1}{2}\sum\limits_{i,j=0}^{3}{{{c}_{ij}}k_{i}^{(1)}}({{\sigma }_{j}}*{{K}^{(2)}}), \\ 
 & =\sum\limits_{i,j=0}^{3}{{{c}_{ij}}k_{i}^{(1)}k_{j}^{(2)}}.  
\end{split}
\ee

Physically, if one considers Hermitian kernels (real coefficients $k_n^{(1)}$ and $k_n^{(2)}$ in (\ref{kernels1})), together with a two-step scanning with a stride equal to the dimension of the respective subsystem, equation (\ref{k2out}) shows that the kernels act as observable operators and the final output represents the expectation value of the corresponding observable \cite{efficient observable operators}. Using the temporary notation $K^A\equiv K^{(1)}$ and $K^B\equiv K^{(2)}$, one can easily check that the output of the first and second convolutional layers are respectively given by $\rho'_{B}=Tr_A \left[ \rho (K^A\otimes I^B) \right]$ and $O=Tr \left[ \rho (K^A\otimes K^B) \right]=\langle K^A\otimes K^B \rangle$.
This method can be generalized to systems with higher dimensions by adjusting the number of kernels, stride, and the dimensions accordingly. For the special case where the kernels correspond to the conjugation of a single Pauli matrix, i.e.  ${{K}^{(1)}}=\sigma _{i}^{*}$, and ${{K}^{(2)}}=\sigma _{j}^{*}$, the CNN output will be equal to the coefficients $C_{ij}$ of the density matrix (\ref{rho}).\\

\begin{figure}[ht]
	\includegraphics[height=8cm]{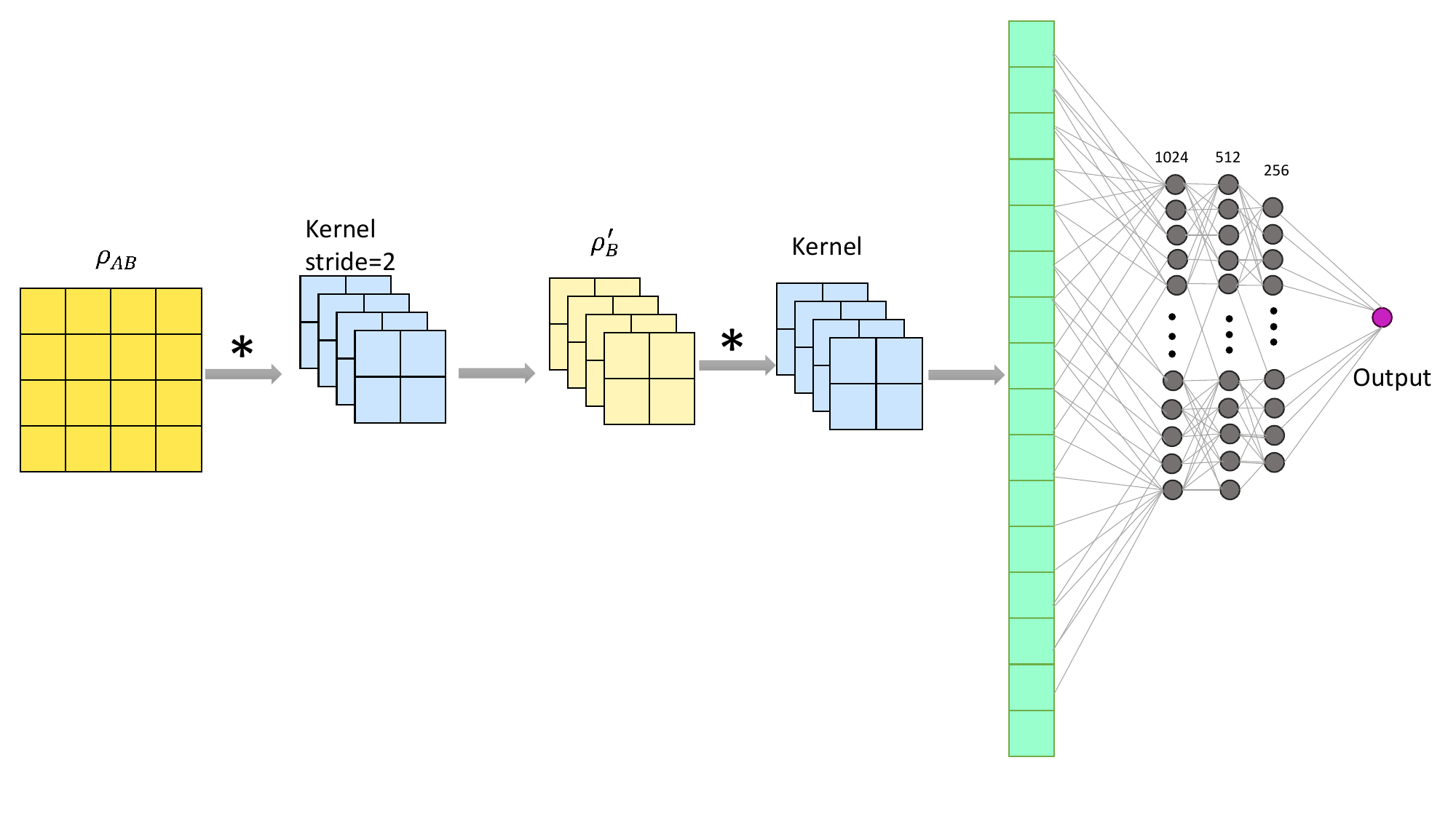}
	\vspace{-0.7cm}
	\centering
	\caption{The structure of our convolutional neural network: The $4\times 4$ input density matrix $\rho_{AB}$ passes through two convolutional layers, each with four $2\times 2$ Hermitian kernels. The first layer has a stride of 2. The output of the second convolutional layer is a vector of length $16$, which passes through three fully connected layers, resulting in a single output. }
	\label{CNN}
\end{figure}

The structure of our network comprises a CNN with two convolutional layers, followed by four fully connected layers which have 1024, 512, 256, and 1 nodes, respectively, see Figure \ref{CNN}.
In the first layer, which has a stride equal to 2, the $4\times 4$ input density matrix is scanned by $4$ Hermitian kernels $\{ K^{(1)}_m,\  m=1,2,3,4 \}$ of dimension $2\times 2$. The output of the first layer, which is a set of four $2\times 2$ matrices, will then be processed within the second layer with four other $2\times 2$ Hermitian kernels $\{ K^{(2)}_n,\  n=1,2,3,4 \}$ . It should be noted that we have not used any activation functions and biases in these two layers. The output of these convolutional layers is a vector of length 16, each element of which is the expectation value of an observable operator as was shown in (\ref{k2out}), i.e.
\be \label{path}
    O_{mn}=Tr\left[ \rho (K^{(1)}_m \otimes K^{(2)}_n)\right], \hspace{1cm} m,n \in \{1,2,3,4\}.
\ee
The extracted feature $O_{mn}$, is formally known as the output of the convolutional path $(m,n)$.  These expectation values are used as the input of the fully connected layers. The activation function of the first three fully connected layers is Relu, and for the final layer, the Sigmoid activation function is used. 
All eight kernel matrices $K^{(1)}_m$ and $K^{(2)}_n$, together with the weights of fully connected layers, are simultaneously optimized to gain the best efficiency for the quantum discord decision problem. It should be noted here that, for a Hermitian input state, if the kernels are initialized to be Hermitian matrices, then their gradient will also be Hermitian. Hence, training the network based on the gradient descent method will leave the Hermiticity of the kernels intact \cite{efficient observable operators}.\\ 

\begin{figure}[H]
	\vspace{-0.5cm}
	\includegraphics[height=4cm]{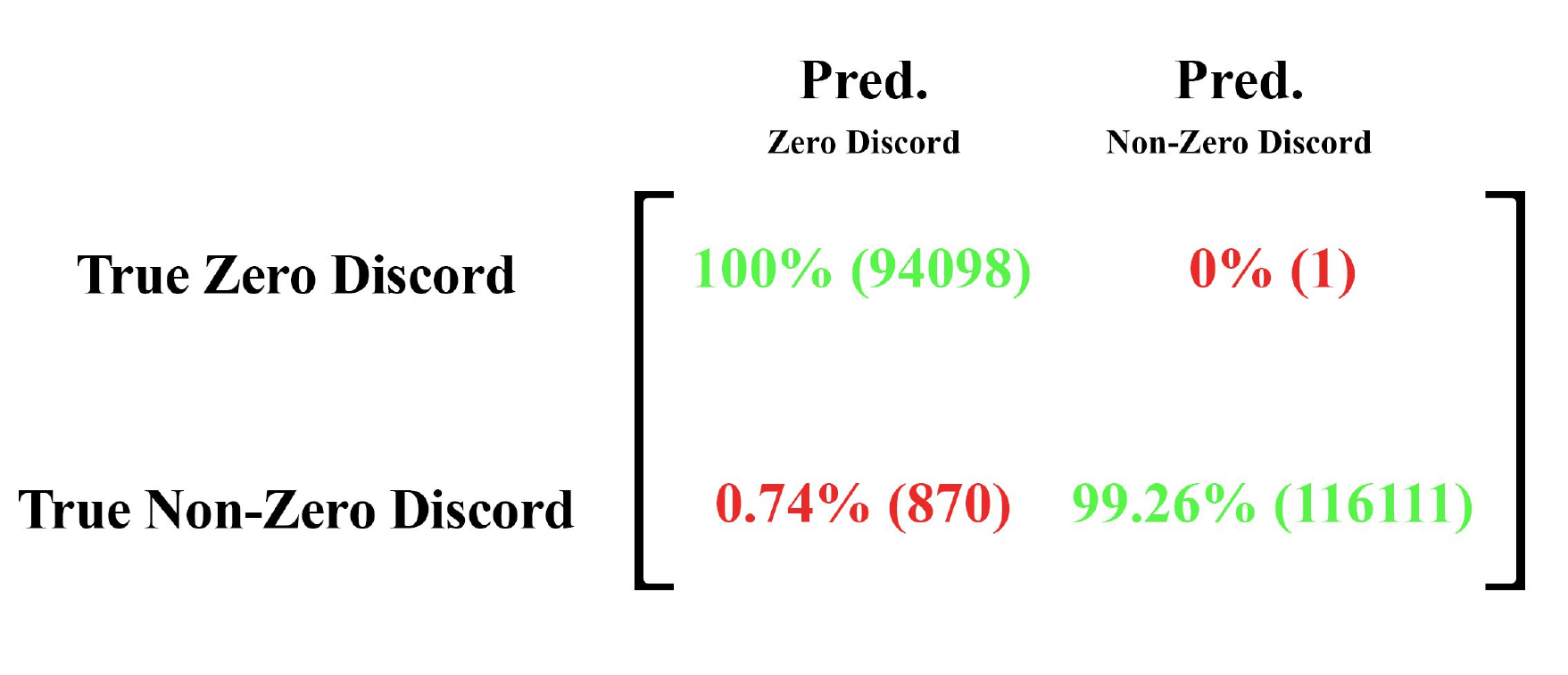}
	\centering
	\vspace{-0.5cm}
	\caption{The confusion matrix of CNN. }
	\label{confusionM}
\end{figure}

We have used a data set of $1055400$ random density matrices of two-qubits, half of which are generated according to equation (\ref{non-discord}). $20$ and $10$ percent of the data are used for the test and validation sets, respectively.
Figure \ref{confusionM} shows the confusion matrix of the optimized  CNN network. The accuracy of this network for the detection of non-discordant states is $100\%$. This perfect performance was expected since the two convolutional layers extract $16$ efficient features of the density matrix, and it is nothing but the complete tomography of the input state, which has $15$ independent parameters. To decrease the number of employed features, we use a branching convolution neural network, which is described below. \\

The superiority of BCNN to CNN is regarded to the fact that BCNN employs fewer kernels, and hence, it is trained faster and consumes fewer computational resources. In a BCNN, one restricts the number of convolutional paths; for example, some of the paths $(m,n)$ in our previous network are selected. The selected non-zero kernels will then be optimized simultaneously to gain the best accuracy. This is similar to the situation if we had turned off some of the kernels of the CNN manually and then optimized the others. \\

\begin{figure}[H]
	\includegraphics[scale=0.55]{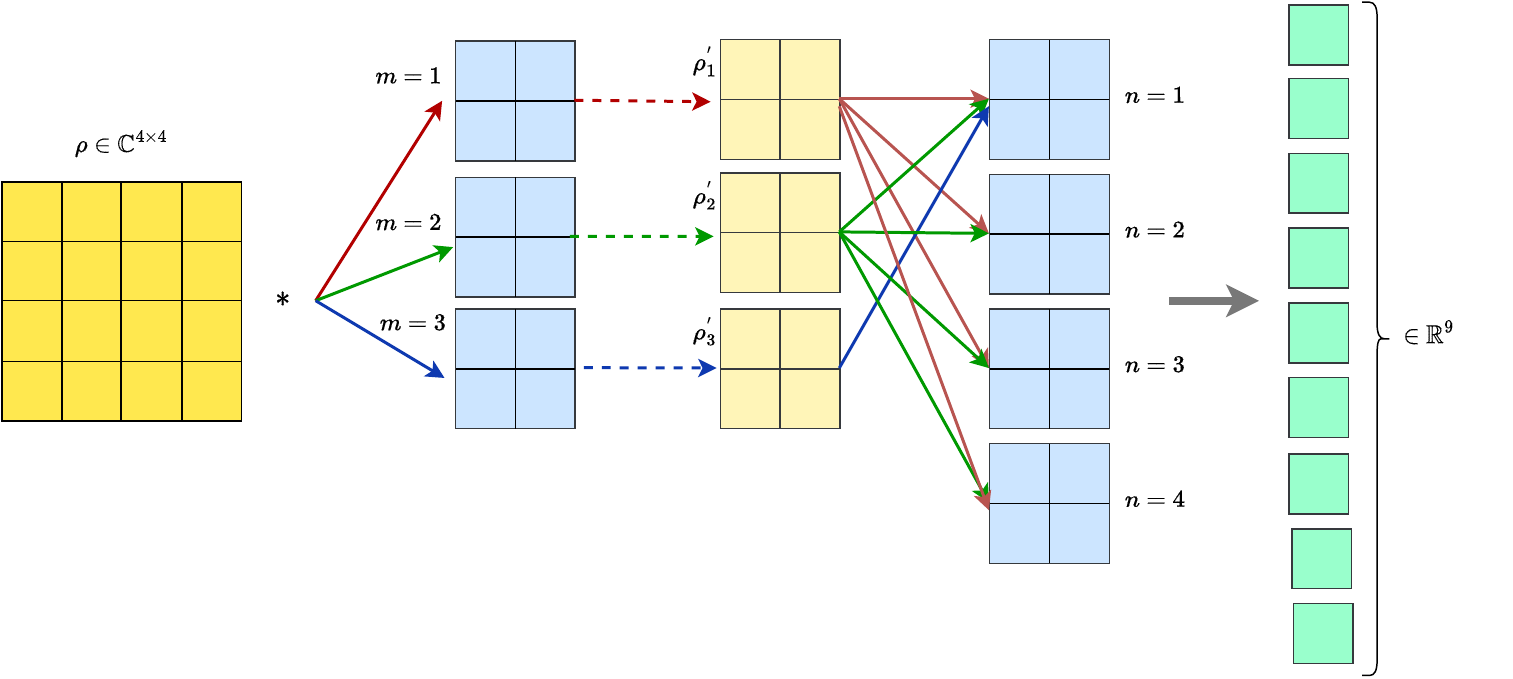}
	\centering
	\vspace{0 cm}
	\caption{Illustration of the method of path selection for BCNN: To increase the number of independent paths $l$, from $2$ to $15$, we fix $m$ and increase $n$ from $1$ to $4$, then we increase $m$ and repeat the process. The arrows show all convolution paths for $l=9$, i.e. the paths $\{(m,n)\}_{l=9}=\{(1,1),\ (1,2),\ (1,3),\ (1,4),\ (2,1), \ (2,2),\ (2,3),\ (2,4),\ (3,1) \}$ }
	\label{introduce BCNN}
\end{figure}

The structure of our BCNN network is exactly the same as our CNN network, i.e. the input states are  $4\times 4$ density matrices, two convolutions layers are followed by four fully connected layers, and all the $2\times 2$ kernels are set to be Hermitian.
The difference is that, here we restrict the number of convolutional paths, and by using a simple pattern, we change the \textit{number of independent convolutional paths}, denoted as $l$, from $2$ to $15$. Similar to the CNN, each convolutional path of our BCNN has a label $(m,n)$ and uses the kernel $K^{(1)}_m \otimes K^{(2)}_n$. Our simple pattern for increasing $l$ is that we fix $m$ and change the value of $n$ from $1$ to $4$, then we increase $m$ and repeat the process. For example, for $l=3$, we select the paths $\{(m,n)\}_{l=3}=\{(1,1),\ (1,2), (1,3)\}$, and for $l=5,6$, we use the paths $\{(m,n)\}_{l=5}=\{(1,1),\ (1,2),\ (1,3),\ (1,4),\ (2,1)\}$ and $\{(m,n)\}_{l=6}=\{(1,1),\ (1,2),\ (1,3),\ (1,4),\ (2,1), \ (2,2)\}$ respectively, see Figure \ref{introduce BCNN}. It is evident that consideration of $16$ convolutional paths is equivalent to considering the CNN network. Note that the output of each convolution path is the single scalar $O_{mn}$, hence the output of a BCNN with $l$ independent paths is a set of $l$ numbers. \\

It is valuable to note that the number of non-trivial kernels is equal in some cases; for example for $l=5$ and $6$, they are explicitly $\{K^{(1)}_1,K^{(1)}_2, K^{(2)}_1, K^{(2)}_2, K^{(2)}_3, K^{(2)}_4 \}$, but the values of the kernels are not the same since the number of convolutional paths is different, and hence the optimization process leads to different results for the kernel matrices. \\



The $l$ real scalar outputs of our BCNN are integrated into a vector, which then passes through a series of four fully connected layers, again with $1024$, $512$, $256$, and $1$ nodes, respectively. The layers include batch normalization, the first three layers use the ReLU activation function, and the final fully connected layer employs a Sigmoid activation function, culminating in the computation of the quantum discord probability. Also, the learning rate has been reduced by a factor of $0.5$ after every $6$ epochs, and these layers optimize the learning and mitigate over-fitting.

\section{Result and Discussion}\label{result}

\begin{figure}[H]
	\vspace{-0.5cm}
	\centering
	\includegraphics[height=8cm]{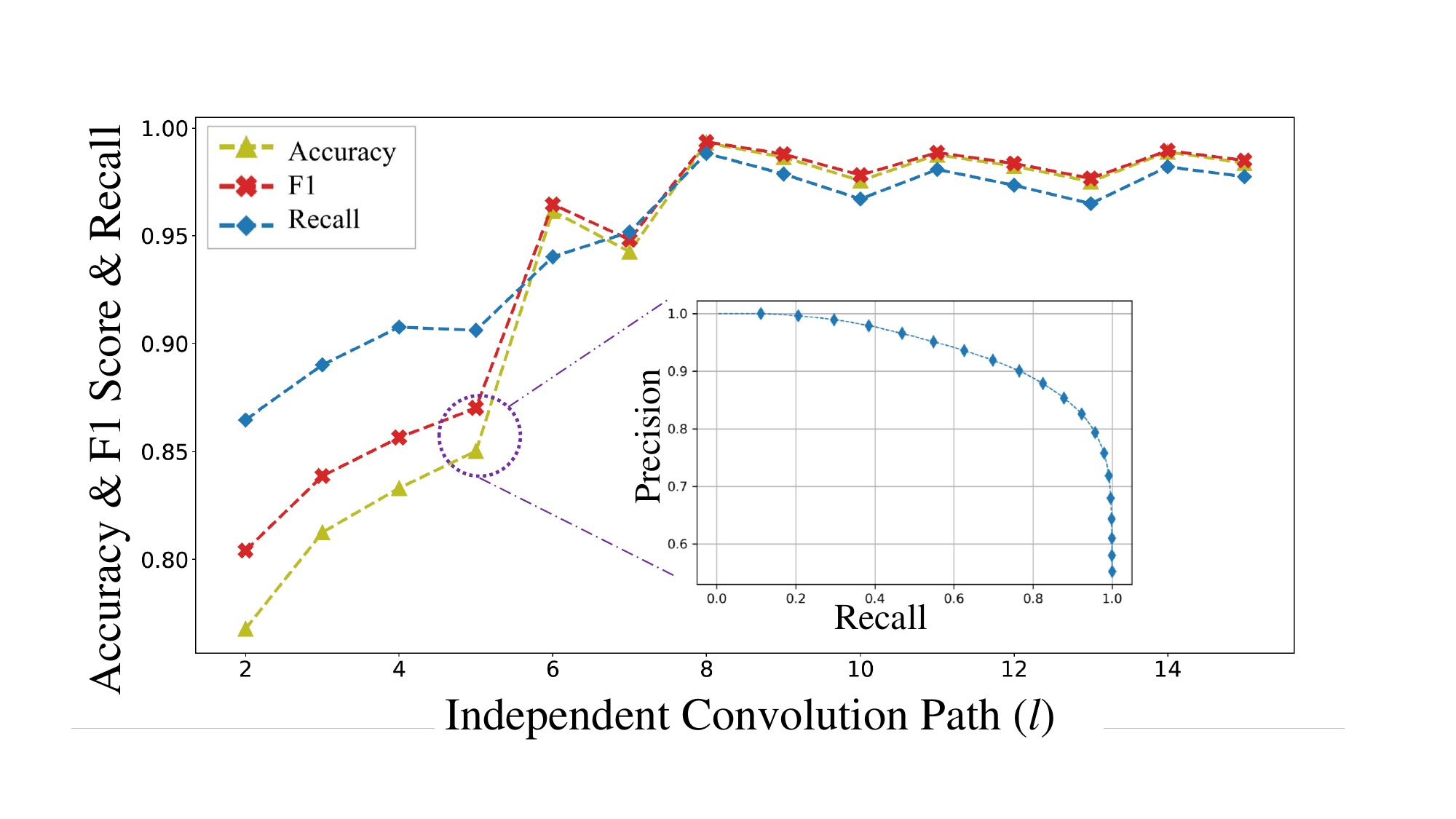}
	\vspace{-0.8cm}
	\caption{Performance of the BCNN network for the quantum discord decision problem of two-qubit states: Accuracy, recall, and F$1$ Score of the model are plotted with respect to the number of convolution paths $l$. By increasing the number of paths, an overall increase in accuracy and recall is seen. }
	\label{BCNNplot}
\end{figure}

To learn our BCNN network of Figure \ref{introduce BCNN}, we again use our data set of $1055400$ random density matrices of two-qubits, half of which are generated according to equation (\ref{non-discord}). $20$ and $10$ percent of the data are respectively used for the test and validation sets.
The model has been trained  for different numbers of convolution paths $2\leq l \leq 15$, and the result is presented in Figure \ref{BCNNplot}. It is evident that one can only use  $8$ paths to reach an accuracy of around $99\%$ for the quantum discord decision problem. In fact, the whole network acts as a quantum discord witness, which only utilizes $8$ efficient features of a quantum state to decide whether it has discord or not. The proposed model also demonstrates an impressive accuracy of about $85\%$ by employing just five independent convolutional paths.
\\

The optimal kernels can  effectively be regarded  as observable operators that are used for the detection of quantum discord.
These trained observable operators (kernels)  are  presented in Table \ref{kernels5,8}, for $5$ and $8$ paths. For the sake of brevity, we have used the expansion (\ref{kernels1}) to  write the coefficient factors only. Remind that the five convolution paths of $l=5$ are $\{(m,n)\}_{l=5}=\{(1,1),\ (1,2),\ (1,3),\ (1,4),\ (2,1)\}$, and one can get the set of $8$ independent paths simply by adding three next paths $\{(2,2),\ (2,3),\ (2,4)\}$ to the above set. \\

\begin{table}[H]
\caption{Kernels of BCNN network with 5 and 8 independent paths. The values have been rounded to three digits and show weight factors of Pauli matrices, according to equation (\ref{kernels1}).}
\label{kernels5,8}
\begin{subtable}[c]{0.5\textwidth}
\subcaption{Five independent paths}
\centering
{\renewcommand{\arraystretch}{1.5}
\begin{tabular}{c|cccc|cccc|}
\cline{2-9}
  & \multicolumn{4}{|c|}{{ {$\boldsymbol{K^{(1)}_m}$} }} &\multicolumn{4}{|c|}{{ {$\boldsymbol{K^{(2)}_n}$} }}  \\ \cline{1-9}
\multicolumn{1}{|c|}{ \textbf{\textit{{m,n}}}} & $\sigma_x^{(1)}$ & $\sigma_y^{(1)}$ & $\sigma_z^{(1)}$ & $I^{(1)}$ &  $\sigma_x^{(2)}$ & $\sigma_y^{(2)}$ & $z_n^{(2)}$ & $I^{(2)}$\\ 
\hline
 \multicolumn{1}{|c|}{\bf 1} & -0.008 & 0.00 & 1.401 & 0.002  & 0.120 & 0.089 & -1.117 & -0.778 \\
 \cline{1-9}
 \multicolumn{1}{|c|}{ \bf 2} & 1.265 & 0.264 & -0.456 & 0.010  &  -0.281 & 0.652 & 0.467 & 0.636  \\ 
 \cline{1-9}
\multicolumn{1}{|c|}{\bf 3} &  &  &  &  & 0.437 & 0.093 & 0.651 & 0.437  \\
\cline{1-9}
\multicolumn{1}{|c|}{ \bf 4} &   &  &  &  & -0.070 & 0.023 & 0.906 & -0.051  \\ 
\hline
\multicolumn{9}{c}{} \\ 
\end{tabular}}
\end{subtable}
\begin{subtable}[c]{0.4\textwidth}
\subcaption{Eight independent paths}
\centering
{\renewcommand{\arraystretch}{1.5}
\begin{tabular}{c|cccc|cccc|}
\cline{2-9}
  & \multicolumn{4}{|c|}{{ {$\boldsymbol{K^{(1)}_m}$} }} &\multicolumn{4}{|c|}{{ {$\boldsymbol{K^{(2)}_n}$} }}  \\ \cline{1-9}
\multicolumn{1}{|c|}{\textbf{\textit{m,n}}} & $\sigma_x^{(1)}$ & $\sigma_y^{(1)}$ & $\sigma_z^{(1)}$ & $I^{(1)}$ &  $\sigma_x^{(2)}$ & $\sigma_y^{(2)}$ & $z_n^{(2)}$ & $I^{(2)}$\\  
\hline
 \multicolumn{1}{|c|}{\bf 1} &   -0.019 & 0.145 & 0.732 & 0.000 & -0.508 & -0.963 & -0.705 & 0.600  \\
 \cline{1-9}
 \multicolumn{1}{|c|}{ \bf 2} & -0.092 & -0.615 & 0.144 & -0.014 & -1.439 & -1.328 & 0.3897 & -0.327 \\ 
 \cline{1-9}
\multicolumn{1}{|c|}{\bf 3} &  &  &  &  & 0.114 & -0.143 & -0.544 & -0.861  \\
\cline{1-9}
\multicolumn{1}{|c|}{ \bf 4} &   &  &  &  & 1.130 & -1.165 & 0.268 & 0.122  \\ 
\hline
\multicolumn{9}{c}{} \\ 
\end{tabular}}
\end{subtable}
\end{table}

Inspired by the above kernels, we will discuss how to design a quantum circuit in order to perform the minimal number of measurements that are necessary for a two-qubit quantum discord decision problem ($5$ or $8$, with regard to our desired accuracy.). 
Since the kernels $K_m^{(i)}$ are considered as observable operators, we omit the weight factor of identity and renormalize other factors to gain new kernels of the form $\tilde{K}_m^{(i)}=\vec{\sigma}^{(i)}.\hat{r}^{(i)}_m$, where $\vec{\sigma}^{(i)}=(\sigma_x^{(i)}, \sigma_y^{(i)}, \sigma_z^{(i)})$, and $\hat{r}_m^{(i)}$ are real unit vectors. The normalized vectors $\hat{r}_m^{(i)}=(x_m^{(i)},y_m^{(i)},z_m^{(i)})$ are presented in Table \ref{normalizedKernels} for each kernel (Note that the sign of coefficient factor of $\sigma_y$ has changed since $\sigma_y^*=-\sigma_y$ ). Then we again optimize our BCNN network of Figure \ref{CNN} with these fixed kernels, and we get the accuracies  $83\%$ and $98\%$, for $5$ and $8$ independent paths, respectively. Note that the expectation values of these new fixed kernels are input values of the fully connected layers of our circuit, and it means that we can replace the branching part of the network with a simple quantum circuit, i.e. we use a quantum circuit to separately measure the expectation values of the normalized fixed kernels, and we then insert these values into our fully connected neural network to decide whether a two-qubit state has quantum discord or not.


\begin{table}[H] 
\caption{Renormalized kernels for BCNN with 5 and 8 independent paths. The table shows the coefficient factors of the expansions $\tilde{K}=\vec{\sigma}.\hat{r}$, and the values are rounded up to three digits.}
\begin{subtable}[c]{0.5\textwidth}
\subcaption{Five independent paths}
\centering
{\renewcommand{\arraystretch}{1.5}
\begin{tabular}{c|ccc|ccc|}
\cline{2-7}
  & \multicolumn{3}{|c|}{{ {$\boldsymbol{\tilde{K}^{(1)}_m}$} }} &\multicolumn{3}{|c|}{{ {$\boldsymbol{\tilde{K}^{(2)}_n}$} }}  \\ \cline{1-7}
\multicolumn{1}{|c|}{\diagbox{ \ \  \textbf{\textit{{m,n}}}}{\ }} & $x_m^{(1)}$ & $y_m^{(1)}$ & $z_m^{(1)}$ &  $x_n^{(2)}$ & $y_n^{(2)}$ & $z_n^{(2)}$ \\ 
\hline
 \multicolumn{1}{|c|}{\bf 1} & 0 &  0 & 1 & 0.106 & -0.079 & -0.991 \\
 \cline{1-7}
 \multicolumn{1}{|c|}{ \bf 2} & \ 0.927 \  & \ -0.193 \  &\  -0.334 \  &\  -0.331 \ & \  -0.767 \  & \ 0.549 \  \\ 
 \cline{1-7}
\multicolumn{1}{|c|}{\bf 3} &  &  & & 0.544 & -0.116 & 0.810  \\
\cline{1-7}
\multicolumn{1}{|c|}{ \bf 4} &  &  &  & -0.077 & -0.025 & 0.997  \\ 
\hline
\multicolumn{7}{c}{} \\ 
\end{tabular}}
\end{subtable}
\begin{subtable}[c]{0.5\textwidth}
\subcaption{Eight independent paths}
\centering
{\renewcommand{\arraystretch}{1.5}
\begin{tabular}{c|ccc|ccc|}
\cline{2-7}
  & \multicolumn{3}{|c|}{{ {$\boldsymbol{\tilde{K}^{(1)}_m}$} }} &\multicolumn{3}{|c|}{{ {$\boldsymbol{\tilde{K}^{(2)}_n}$} }}  \\ \cline{1-7}
\multicolumn{1}{|c|}{\diagbox{\ \  \textbf{\textit{m,n}}}{\ }} & $x_m^{(1)}$ & $y_m^{(1)}$ & $z_m^{(1)}$ &  $x_n^{(2)}$ & $y_n^{(2)}$ & $z_n^{(2)}$ \\  
\hline
 \multicolumn{1}{|c|}{\bf 1} & \  -0.025 \ & \  0.194 \ & \ 0.981 \  & \  -0.392 \ & \ 0.742 \  & \ -0.543 \ \\
 \cline{1-7}
 \multicolumn{1}{|c|}{ \bf 2} & -0.144 &-0.964 & 0.226 &-0.721 &0.665 & 0.195 \\ 
 \cline{1-7}
\multicolumn{1}{|c|}{\bf 3} &  &  &  &0.199 &0.249 & -0.948  \\
\cline{1-7}
\multicolumn{1}{|c|}{ \bf 4} &  &  &  & 0.687 & 0.708 & 0.163  \\ 
\hline
\multicolumn{7}{c}{} \\ 
\end{tabular}}
\end{subtable}
\label{normalizedKernels}
\end{table}

\begin{figure}[h]
    \centering
    \includegraphics[scale=0.6]{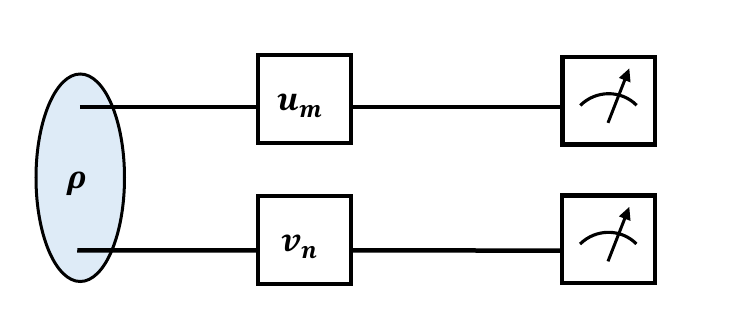}
    \vspace{-0.5 cm}
    \caption{Quantum circuit for extracting the feature $\tilde{O}_{mn}$. }
    \label{measure}
\end{figure}

To design the circuit, we use the fact that any arbitrary trace-less $2\times 2$ operator can be converted to the Pauli $\sigma_z$ operator by applying a suitable unitary matrix. This leads to the following relation for the expectation values of the renormalized kernels,
\begin{eqnarray}
    \tilde{O}_{mn}:=Tr\left[\rho (\tilde{K}^{(1)}_m \otimes \tilde{K}^{(2)}_n)\right]&=& Tr\left[\rho \left((\vec{\sigma}^{(1)}.\hat{r}^{(1)}_m) \otimes (\vec{\sigma}^{(2)}.\hat{r}^{(2)}_n)\right)\right] \cr
    &=& Tr\left[\rho \left( (u_m^\dagger \otimes v_n^\dagger) (\sigma_z^{(1)} \otimes \sigma_z^{(2)}) (u_m \otimes v_n^) \right)\right] \cr
    &=& Tr\left[  \left( (u_m \otimes v_n ) \rho (u_m^\dagger \otimes v_n^\dagger) \right) (\sigma_z^{(1)} \otimes \sigma_z^{(2)})  \right],
\end{eqnarray}
where $u_m$ and $v_n$ are unitary operators that are chosen such that $u_m \ \vec{\sigma}^{}.\hat{r}^{}_m \ u_m^\dagger=  \sigma_z^{} $ and $v_n \ \vec{\sigma}^{}.\hat{r}^{}_n \ v_n^\dagger =  \sigma_z^{} $, simply $u_m= |0\rangle \langle r_m^+|+|1\rangle \langle r_m^-|$, where $|r_m^\pm\rangle$ and $|0\rangle$ or $|1\rangle$ are the eigenvectors of $\vec{\sigma}.\hat{r}_m$ and $\sigma_z$ Pauli operator with positive and negative eigenvalues. A similar equation describes $v_n$ too. 
The above equation means that the desired expectation values can be found by applying suitable rotations $u_m$ and $v_n$ on the density matrix, followed by simultaneously $\sigma_z$ measurements of both qubits, see Figure \ref{measure}. The number of different required $u_m \otimes v_n$ unitaries is equal to the number of different paths of the BCNN network, i.e. $5$ or $8$ in our case. For the case of $5$ measurements, one only needs to use the pairs $\{ u_1 \otimes v_1, u_1 \otimes v_2, u_1 \otimes v_3, u_1 \otimes v_4, u_2 \otimes v_1 \} $.

\section{Conclusions}\label{con.}
We have investigated the problem of quantum discord detection for general two-qubit states by using convolutional neural networks. We have shown that it is possible to precisely classify discordant and non-discordant states by using a CNN network with two convolutional layers and $16$ Hermitian kernels. We have discussed that the Hermitian kernels can be regarded as observable operators, and using $16$ kernels for efficient feature extraction is equivalent to performing $16$ efficient measurements, which is nothing but the complete tomography of the state. 
We have also designed a branching convolutional neural network that only uses $5$ or $8$ kernels to attain an accuracy of $85\%$ or $99\%$,  respectively, for the same problem. This result contributes to the fundamental understanding of quantum discord by shedding light on the observable operators (kernels) that are necessary for discord detection. By renormalizing the optimized kernels of our BCNN network, we have proposed a quantum circuit, the expectation values obtained from which  
will be inserted in a fully connected network in order to decide whether a quantum state has discord or not. This network attains an accuracy of $83\%$ or $98\%$ for $5$ or $8$ normalized kernels. The fully connected network is then similar to a discord witness that takes the efficient features (from the convolutional layers or alternatively from the measurement results) and decides on the existence of quantum discord. 
The results obtained here highlight the considerable potential of BCNN for quantum discord detection of two-qubit states. The usefulness of BCNN for discord detection of higher dimensional states can be investigated in future works.

{}
\end{document}